\documentclass[sn-mathphys-num]{sn-jnl}

\usepackage{graphicx}%
\usepackage{multirow}%
\usepackage{amsmath,amssymb,amsfonts}%
\usepackage{amsthm}%
\usepackage{mathrsfs}%
\usepackage[title]{appendix}%
\usepackage{xcolor}%
\usepackage{textcomp}%
\usepackage{manyfoot}%
\usepackage{booktabs}%
\usepackage{algorithm}%
\usepackage{algorithmicx}%
\usepackage{algpseudocode}%
\usepackage{listings}%
\usepackage{subcaption}
\usepackage{siunitx}
\usepackage{empheq}

\theoremstyle{thmstyleone}%
\theoremstyle{thmstyletwo}%

\theoremstyle{thmstylethree}%

\begin{document}

\title[Article Title]{Single-Shot, Single-Mask X-ray Dark-field and Phase Contrast Imaging}

\author[1]{\fnm{Jingcheng} \sur{Yuan}}\email{jyuan10@uh.edu}

\author*[1,2,3]{\fnm{Mini} \sur{Das}}\email{mdas@uh.edu}

\affil[1]{\orgdiv{Department of Physics}, \orgname{University of Houston}, \orgaddress{\street{3507 Cullen Blvd}, \city{Houston}, \postcode{77204}, \state{Texas}, \country{USA}}}

\affil[2]{\orgdiv{Department of Electrical and Computer Engineering}, \orgname{University of Houston}, \orgaddress{\street{4226 Martin Luther King Blvd}, \city{Houston}, \postcode{77204}, \state{Texas}, \country{USA}}}

\affil[3]{\orgdiv{Department of Biomedical Engineering}, \orgname{University of Houston}, \orgaddress{\street{3517 Cullen Blvd}, \city{Houston}, \postcode{77204}, \state{Texas}, \country{USA}}}


\abstract{X-ray imaging, traditionally relying on attenuation contrast, struggles to differentiate materials with similar attenuation coefficients like soft tissues. X-ray phase contrast imaging (XPCI) and dark-field (DF) imaging provide enhanced contrast by detecting phase shifts and ultra-small-angle X-ray scattering (USAXS). However, they typically require complex and costly setups, along with multiple exposures to retrieve various contrast features. In this study, we introduce a novel single-mask X-ray imaging system design that simultaneously captures attenuation, differential phase contrast (DPC), and dark-field images in a single exposure. Most importantly, our proposed system design requires just a single mask alignment with relatively low-resolution detectors. Using our novel light transport models derived for these specific system designs, we show intuitive understanding of contrast formation and retrieval method of different contrast features. Our approach eliminates the need for highly coherent X-ray sources, ultra-high-resolution detectors, spectral detectors or intricate gratings. We propose three variations of the single-mask setup, each optimized for different contrast types, offering flexibility and efficiency in a variety of applications. The versatility of this single-mask approach along with the use of befitting light transport models holds promise for broader use in clinical diagnostics and industrial inspection, making advanced X-ray imaging more accessible and cost-effective. }

\keywords{X-ray imaging, phase contrast, dark-field imaging, single-mask set-up, X-ray wave optics, X-ray scattering}



\maketitle

\section{Introduction}\label{sec1}
X-ray imaging has long been a fundamental tool in both medical diagnostics and materials science due to its ability to reveal internal structures with high spatial resolution. Traditional X-ray imaging techniques predominantly rely on attenuation contrast, where differences in the absorption of X-rays by various tissue types or materials generate an image. However, this method often falls short when distinguishing between materials with similar attenuation coefficients, especially for light-element materials like soft tissues. Additionally, its ability to detect microstructures is constrained by the resolution limits of conventional imaging systems.

X-ray phase contrast imaging (XPCI) and dark-field (DF) imaging have emerged as powerful complementary techniques that offer enhanced contrast, particularly for soft tissues and materials with microstructures \cite{fitzgeraldPhaseSensitiveRay2000,lewisMedicalPhaseContrast2004,auweterXrayPhasecontrastImaging2014,taoPrinciplesDifferentXray2021,quenotXrayPhaseContrast2022}. Phase contrast imaging is sensitive to the phase shift of the X-ray wave that passes through the medium, which exploits variations in the refractive index of different materials at the length scale of tens of micrometers. These variations induce beam refraction at the micro-radian scale \cite{taoPrinciplesDifferentXray2021,quenotXrayPhaseContrast2022}. This type of techniques offers enhanced visualization through signatures like differential phase contrast (DPC) and Laplacian phase signatures, etc.

Dark-field imaging, on the other hand, is sensitive to ultra-small-angle X-ray scattering (USAXS) induced by sub-pixel-scale (microns or sub-micron level) microstructures within the medium. As X-rays pass through, they undergo multiple refractions caused by these microstructures, resulting in USAXS within an extremely narrow angular range, typically at the micro-radian scale \cite{quenotXrayPhaseContrast2022}. This scattering manifests as localized blurring in the images. This phenomenon highlights the presence of microstructures that are invisible in traditional absorption-based imaging, thereby introducing a new dimension of contrast in X-ray imaging.

X-ray refraction and ultra-small-angle scattering are both related to spatial variations of refractive index within the medium, but on different length scales: refraction is caused by the variations that are resolvable by x-ray detectors (typically tens of microns pitch) while small-angle scattering is induced by unresolved microstructures at a sub-pixel scale. \cite{lynchInterpretationDarkfieldContrast2011}

Over the past few decades, several approaches have been developed to achieve X-ray phase contrast and dark-field imaging, each offering unique advantages but also presenting significant challenges hindering clinical translation. The detection of X-ray refraction and USAXS angles, which are typically in the micro-radian range, remains particularly difficult.

Propagation-based (PB) imaging  \cite{wilkinsPhasecontrastImagingUsing1996,suzukiXrayRefractionenhancedImaging2002} is the simplest technique for phase contrast imaging, as it does not require additional optical elements in the beam path. The method works by increasing the sample-to-detector distance, allowing phase contrast patterns to form as the propagation of X-rays. Despite its simplicity, PB imaging requires a coherent or partially coherent X-ray source and longer exposure times to produce high-quality phase-contrast images. Also, retrieving the phase signal from thick, heterogenous and complex samples can be challenging, as it usually has extra requirements, like single-material assumption \cite{paganinSimultaneousPhaseAmplitude2002}, multiple exposures \cite{yanAttenuationpartitionBasedIterative2008}, or spectral imaging \cite{gursoySinglestepAbsorptionPhase2013}, etc.

Grating interferometry  \cite{weitkampXrayPhaseImaging2005,pfeifferPhaseRetrievalDifferential2006} typically employs three precise gratings to leverage the Talbot effect in extracting phase contrast and dark-field signals. This method stands out as the only phase contrast technique that has been successfully scaled for human-sized prototypes \cite{frankDarkfieldChestXray2022,viermetzDarkfieldComputedTomography2022}. However, its clinical translation and application faces challenges due to the complexity of the setup, stringent fabrication requirements for gratings with micron-level pitch, and the need for multiple exposures to retrieve differential phase and dark-field image. Furthermore, the technique is typically optimized for a single wavelength, which limits its performance with broad-spectrum X-ray sources.

Modulation-based (or speckle-based) imaging \cite{morganXrayPhaseImaging2012,pil-aliMultimodalSingleshotXray2023} and mesh-based imaging  \cite{petruccelliPhaseCoherentScatter2016} simplify phase contrast setups by using a single optical element, such as a periodic mask or random pattern like sandpaper \cite{beltranFastImplicitDiffusive2023}. While these methods offer simplicity by eliminating the need for precise alignment, they rely heavily on ultra-high-resolution detectors to capture the micron-scale distortions in the reference pattern caused by the small refraction and scattering angles induced by the object. This dependence on high detector resolution, along with the requirement for highly spatially coherent X-ray sources, remains a major barrier to their translation in clinical and industrial settings.

The edge-illumination (or coded-aperture) method  \cite{olivoCodedapertureTechniqueAllowing2007} uses two lower-pitch masks instead of delicate gratings to capture phase contrast signal with a simpler set-up. Despite its advantages, it requires precise alignment between the two masks and the detector, and highly spatially coherent X-ray sources. Also, it requires multiple exposures \cite{olivoEdgeilluminationXrayPhasecontrast2021} or spectral imaging \cite{dasSpectralXrayPhase2014} to extract dark-field and differential phase images.

In summary, most of these techniques demand complex and costly setups, often requiring one or more of the following: highly coherent X-ray sources, ultra-high-resolution detectors, sophisticated X-ray gratings, ultra-precise alignment of components, and multiple exposures to retrieve phase contrast images. These stringent requirements have hindered the broader adoption of these methods in clinical and industrial applications.

\begin{figure} [ht]
    \centering
    \begin{subfigure}{0.6\textwidth}
        \centering
        \includegraphics[width=\textwidth]{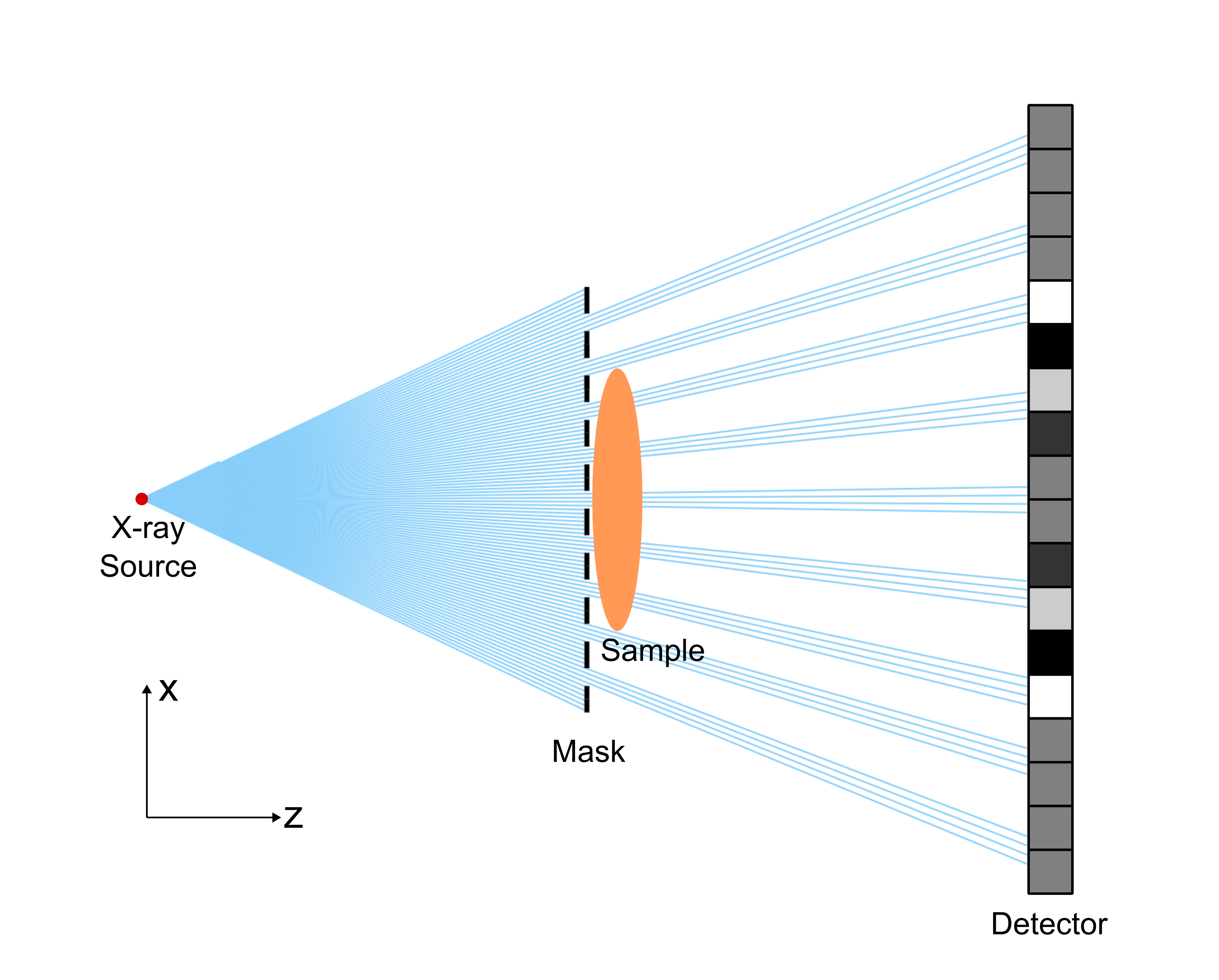}
        \caption{}\label{fig:diagram_single_mask}
    \end{subfigure}
    \begin{subfigure}{0.3\textwidth}
        \centering
        \includegraphics[width=\textwidth]{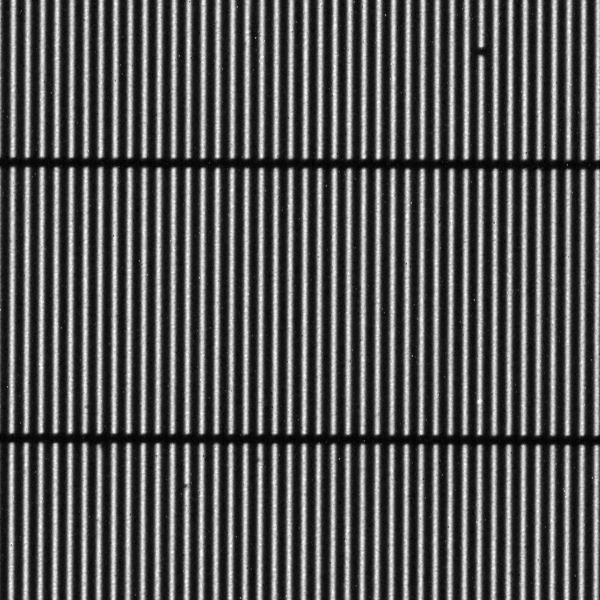}
        \caption{}\label{fig:mask_image}
    \end{subfigure}
\caption{(a) Diagram for single-mask X-ray differential phase contrast imaging system. (b) Zoomed in high-resolution (\SI{8}{\micro\metre}) x-ray image of the mask. The period and opening width of the mask is \SI{53}{\micro\metre} and \SI{20}{\micro\metre}, respectively. The horizontal lines are supporting structures of the mask.}\label{fig:diagram}
\end{figure}

The single mask method, pioneered by Krejci \textit{et al.} \cite{krejciHardXrayPhase2010}, has emerged as a promising approach to x-ray phase contrast imaging. This technique requires only one optical element: a periodic absorption mask positioned between the x-ray source and the sample, typically near the sample itself (Fig. \ref{fig:diagram_single_mask}). The mask creates x-ray beamlets (Fig. \ref{fig:mask_image}) by periodically blocking x-rays with thin strips of heavy-element materials, such as gold. As x-rays pass through the sample, various interactions—attenuation, refraction, and scattering—modify the diffraction pattern formed by the mask on the detector plane. Specifically, x-ray attenuation locally reduces the average brightness of the pattern, refraction causes slight local displacements or distortions, and scattering causes local blurring, thereby decreasing the contrast of the pattern stripes. The fringe displacements caused by refraction are typically extremely small, often at the micron or sub-micron level. However, standard clinical x-ray detectors generally lack the resolution needed to detect such minute displacements accurately. To overcome this limitation, the single-mask method utilizes a specialized alignment between the mask and the detector pixels, with every pixel acting as a collimator, to capture these tiny displacements that would be of sub pixel order with a detector resolution close to clinical usage.

Building on the light-transport model published in our previous work \cite{yuanTransportofintensityModelSinglemask2024}, the single-mask method is capable of simultaneously capturing attenuation and differential phase contrast (DPC) images in a single exposure, offering high signal-to-noise ratio and dose efficiency in both radiography \cite{yuanComparingSNRBenefits2022} and phase contrast micro-CT \cite{yuanSinglemaskPhaseImaging2024}. However, while previous studies have demonstrated its ability to extract attenuation and DPC images, this imaging system has not yet been capable of simultaneously capturing dark-field images.

In this study, we propose three variations of the single-mask setup with the capability to obtain different contrast features. Our practical and translatable methods show that one can simultaneously capture attenuation, differential phase, and dark-field signatures in a single-shot imaging using a relatively low resolution detector when our proposed X-ray mask alignment is used. Retrieving these independent signals would yield quantitative information of the underlying material properties that impact these signal formations. There multiple image features or material properties would also allow fast and effective material classification.

The new physics models we propose here for these novel system designs would allow intuitive understanding of signal formation in these geometries. In addition, it also allows effective retrieval of these independent contrast features from single exposure without the need for spectral information or complex motion of optical elements. These models draw inspiration from our previous work \cite{yuanTransportofintensityModelSinglemask2024} and the Fokker-Planck equation \cite{paganinXrayFokkerPlanck2019}.

\section{Light Transport Model}
In our previous work, our formulation for single-mask DPC configuration is based on the approximated transport-of-intensity equation (TIE) \cite{zuoTransportIntensityEquation2020}, which described the intensity change after the EM wave propagates for a certain distance $z$:
\begin{equation}
    \begin{aligned}
    I(\vec{r},z) &= I(\vec{r},0)-\frac{z}{k}\nabla_\perp\cdot[I(\vec{r},0)\nabla_\perp\phi(\vec{r},0)] \\
    \end{aligned}
\end{equation}
where $I(\vec{r},0)$ is the x-ray intensity before propagation, $\vec{r}=(x,y)$ is the transverse coordinate, and $k$ is the wave number. The second term describes changes in intensity due to variations of phase shift $\phi(\vec{r},0)$ after propagating a distance $z$, which leads to a redistribution of optical energy, including local convergence/divergence and transverse displacement of the beams. The X-ray Fokker-Planck equation, proposed by Paganin \textit{et al.} \cite{paganinXrayFokkerPlanck2019}, is a modification of TIE that adding a term modeling ultra-small-angle scattering:
\begin{equation}
    \begin{aligned}
        I(\vec{r},z) = I(\vec{r},0)-\frac{z}{k}\nabla_\perp\cdot[I(\vec{r},0)\nabla_\perp\phi(\vec{r},0)] + z F(\vec{r})\nabla_\perp^2[D(\vec{r})I(\vec{r},z)]
    \end{aligned}
\end{equation}
where $D(\vec{r})$ is the property related to distribution of scattered photons and $F(\vec{r})$ is the fraction of the optical energy converted to USAXS. If we assume $D(\vec{r})$ is slowly varying within the transverse plane, we get:
\begin{equation}
    \begin{aligned}
        I(\vec{r},z) = I(\vec{r},0)-\frac{z}{k}\nabla_\perp\cdot[I(\vec{r},0)\nabla_\perp\phi(\vec{r},0)] +z F(\vec{r})D(\vec{r})\nabla_\perp^2 I(\vec{r},0)
    \end{aligned}
\end{equation}
Here $F(\vec{r})D(\vec{r})$ is defined as the effective scattering coefficient, which denotes the structural property of the sample:
\begin{equation}
    S(\vec{r}) = F(\vec{r})D(\vec{r})
\end{equation}
When applying to single-mask configuration, $I(\vec{r},0)=T(\vec{r})M(x)$, where $T$ and $M$ is the transmission function of the object and the mask, respectively. Similar to our previous work \cite{yuanTransportofintensityModelSinglemask2024}, the X-ray intensity measured by a certain detector pixel can be written as the integration within the pixel region:
\begin{equation}
    \begin{aligned}
    I_n=\int_{x_n}^{x_{n+1}}I(x,z)\mathrm{d}x
    \end{aligned}
\end{equation}
For simplicity, here we focus on a single detector row and $I_n$ is the intensity measured by the $n^{th}$ pixel of this row. The equation can be derived as:
\begin{equation}
    \begin{aligned}
        I_n &= T(x_n)\int_{x_n}^{x_{n+1}} M(x)\mathrm{d}x - \frac{z}{k}T(x_n)\nabla_\perp^2\phi(x_n)\int_{x_n}^{x_{n+1}}M(x)\mathrm{d}x \\
        &- \frac{z}{k}T(x_n)\partial_x\phi(x_n)\int_{x_n}^{x_{n+1}}M'(x)\mathrm{d}x + z T(x_n)S(x_n) \int_{x_n}^{x_{n+1}}M''(x)\mathrm{d}x
    \end{aligned}\label{eqn:general}
\end{equation}
With this formulation in place, we will now describe various configurations of single-mask phase imaging that offers enhancement and recovery of different signal types. 

\section{Three Configurations of Single-Mask Method}
\subsection{Single-Mask DPC Configuration}
\begin{figure} [ht]
    \centering
    \begin{subfigure}{0.5\textwidth}
        \centering
        \includegraphics[width=\textwidth]{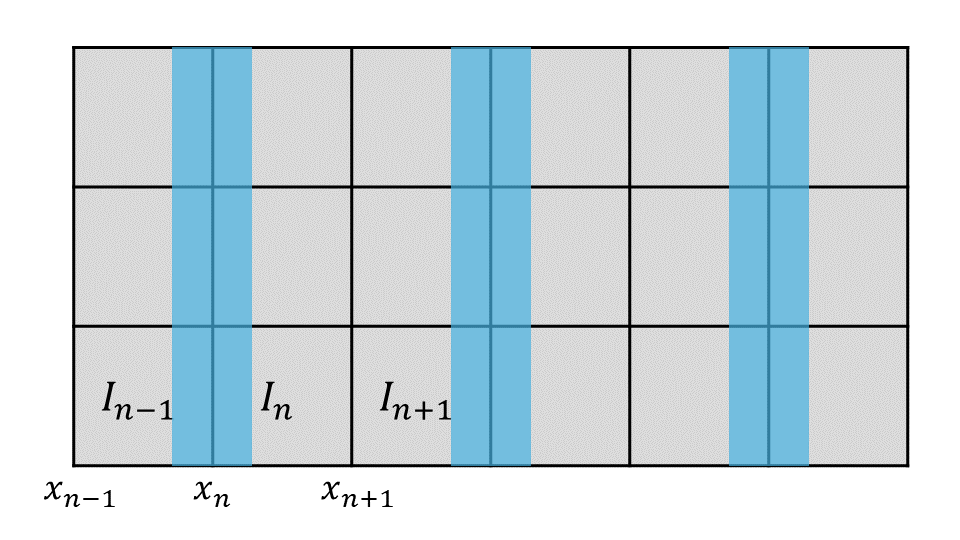}
        \caption{}\label{fig:diagram_DPC}
    \end{subfigure}
    \begin{subfigure}{0.34\textwidth}
        \centering
        \includegraphics[width=\textwidth]{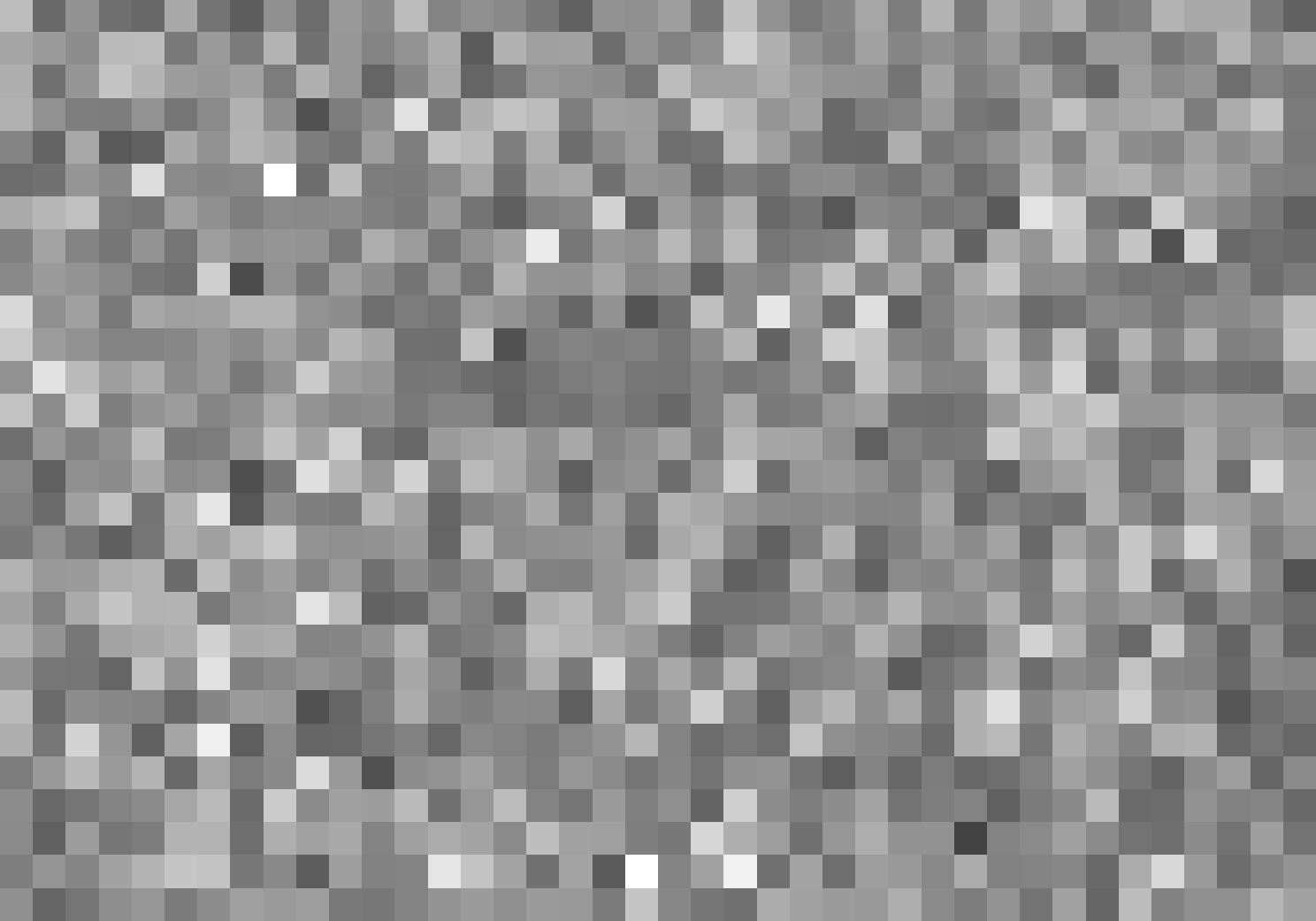}
        \caption{}\label{fig:mask_DPC}
    \end{subfigure}
\caption{(a) Diagram of the single-mask differential phase contrast (DPC) configuration. The blue strips represent the X-ray beamlets created by the mask, while the squares represent the detector pixels, with their color indicating the X-ray intensity measured by each pixel. (b) Zoomed-in mask image from the experiment. The mask pattern is not visible, as each pixel is uniformly illuminated by the X-ray beamlets.}\label{fig:DPC}
\end{figure}

In the original single-mask method \cite{krejciHardXrayPhase2010,yuanTransportofintensityModelSinglemask2024} (referred to as DPC configuration in this article), the mask should be positioned such that the period of the projected mask pattern is twice the detector pixel size $p$, and the center of each beamlet strip is aligned with every other pixel boundary (Fig. \ref{fig:diagram_single_mask} and Fig. \ref{fig:diagram_DPC}). 

Without a sample in the beam path, this specific alignment creates almost uniform illumination of each detector pixel (Fig. \ref{fig:mask_DPC}). When a sample is introduced into the beam path, some beamlets are shifted from their original positions due to refraction by the sample, resulting in intensity variations at the corresponding pixels. Consequently, the refraction angle or differential phase signal can be detected using a low-resolution detector. The intensity received by each pixel can be derived as shown in the following equation.

\begin{equation}
    I_n= w_e T_n(1-L_n) - \alpha (-1)^n T_n D_n 
    \label{eqn:DPC}
\end{equation}
In this equation, $T_n = T(x_n)$, $D_n = \frac{z}{k}\partial_x\phi(x_n)$, and $L_n = \frac{z}{k}\nabla_\perp^2\phi(x_n)$, which represents the average attenuation, differential phase, and Laplacian phase across the corresponding pixel, respectively. 

Also, the two coefficients in the equation, $w_e$ and $\alpha$, are parameters related to the mask, corresponding to its effective transparency and equivalent contrast, respectively. 

From Eqn. (\ref{eqn:DPC}), we can observe the following: without the sample ($T_n=1$, $L_n=0$, and $D_n=0$), the detector produces a uniform image, and the mask pattern is invisible, with the intensity measured by each pixel determined by the effective transparency $w_e$ of the mask. Once the sample is introduced into the beam path, the differential phase signal $D_n$ contributes to the second term, resulting in bright and dark fringes on the image, thereby revealing the mask pattern.

As derived in our previous article \cite{yuanTransportofintensityModelSinglemask2024}, the retrieval method can be written as: 
\begin{equation}
    \left\{
    \begin{aligned}
        T_n(1-L_n) &\approx \frac{\bar{I}_n + \bar{I}_{n+1}}{2} \\
        \frac{\alpha}{w_e}D_n \approx & \frac{\bar{I}_n-\bar{I}_{n+1}}{\bar{I}_n+\bar{I}_{n+1}}(-1)^n
    \end{aligned}\right.
    \label{eqn:DPC_retreival}
\end{equation}
where $\bar{I}_n = I_n^{(S)}/I_n^{(M)}$ is the flat-field corrected intensity, in which $I_n^{(S)}$ and $I_n^{(M)}$ is the image with mask and sample, and image with mask only, respectively. The right-hand side of the two equations can be regarded as retrieved attenuation with Laplacian phase enhancement and DPC image, respectively. Also, we note the intensity of the DPC image equals the differential phase term $D_n$ multiplied by $\alpha/w_e$, which can be understood as the DPC sensitivity of the imaging system.
\begin{figure}[ht]
	\centering
	\begin{subfigure}{0.4\textwidth}
		\includegraphics[width=\textwidth]{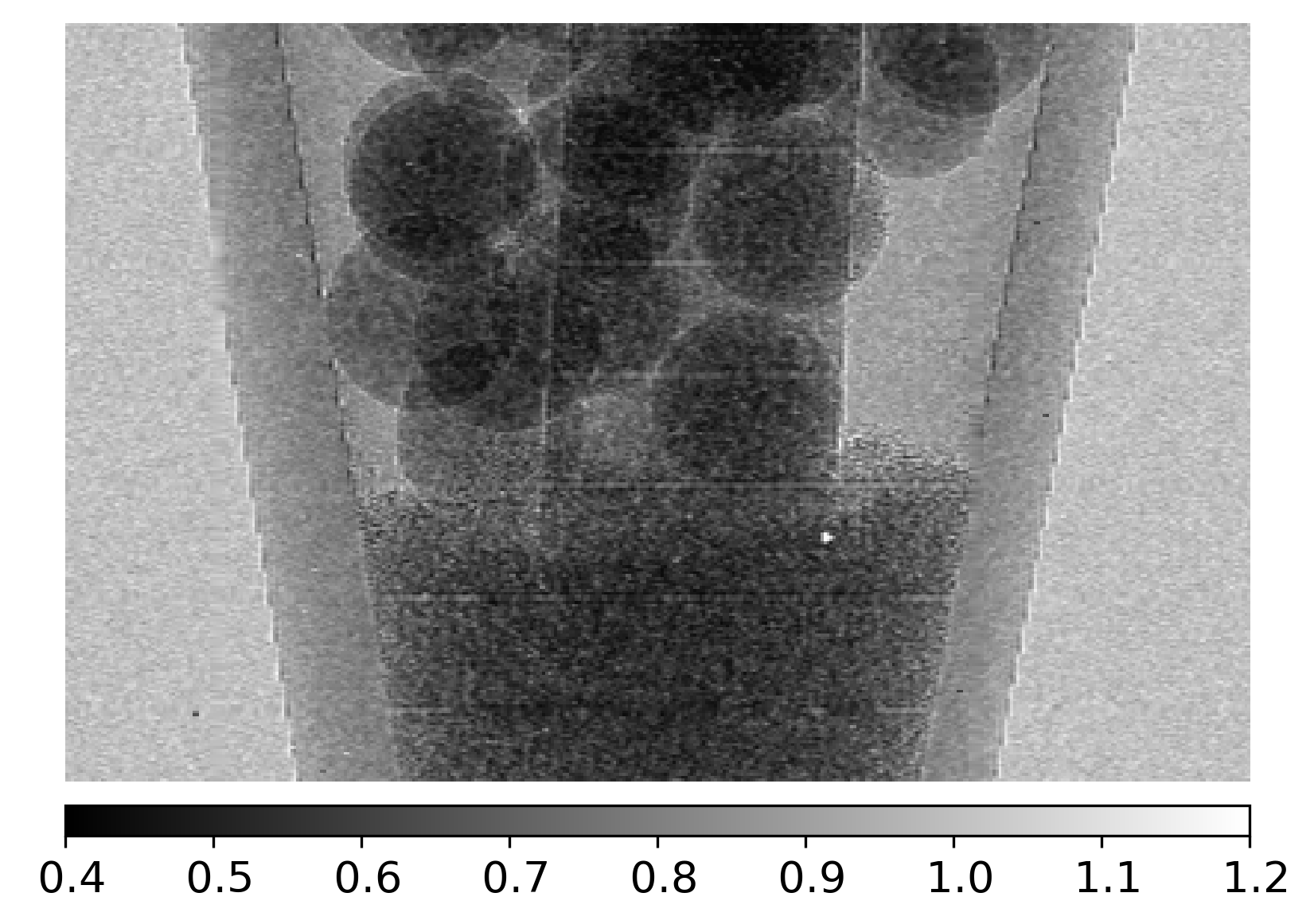}
		\caption{} \label{fig:DPC_T}
	\end{subfigure}
	\begin{subfigure}{0.4\textwidth}
		\includegraphics[width=\textwidth]{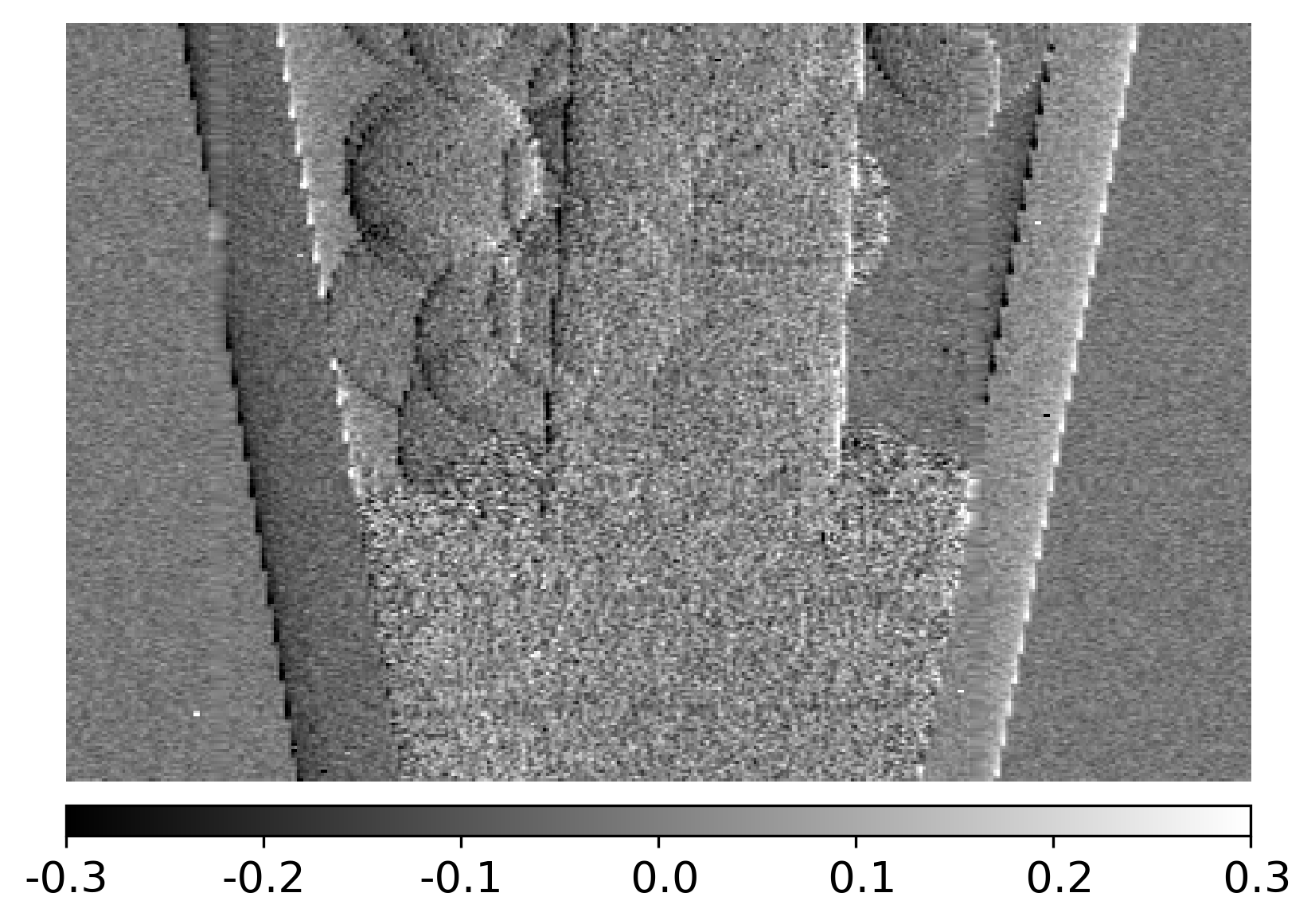}
		\caption{} \label{fig:DPC_D}
	\end{subfigure}
	\caption{Retrieved (a) Attenuation image and (b) DPC image of the multi-material phantom with single mask DPC configuration}\label{fig:phantom_DPC}
\end{figure}

Fig. \ref{fig:phantom_DPC} shows the retrieved attenuation and differential phase images retrieved from a single shot with the single-mask DPC configuration. The sample used was a multi-material phantom, which included a graphite rod with a diameter of 3 mm, several plastic beads with a diameter of 2 mm, and diamond powders with grain size ranging from 50 to \SI{80}{\micro\metre}. All components were placed inside a centrifuge tube. We can see the DPC image demonstrate superior contrast compared to the attenuation image, effectively highlighting the boundaries between various objects.

\subsection{Single Mask Dark-Field Configuration}
\begin{figure} [ht]
    \centering
    \begin{subfigure}{0.5\textwidth}
        \centering
        \includegraphics[width=\textwidth]{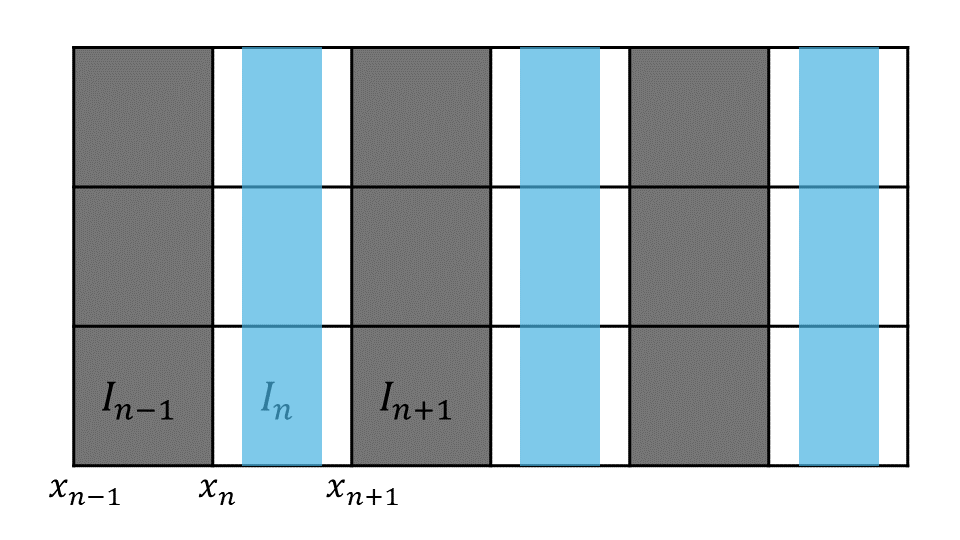}
        \caption{}\label{fig:diagram_DF}
    \end{subfigure}
    \begin{subfigure}{0.34\textwidth}
        \centering
        \includegraphics[width=\textwidth]{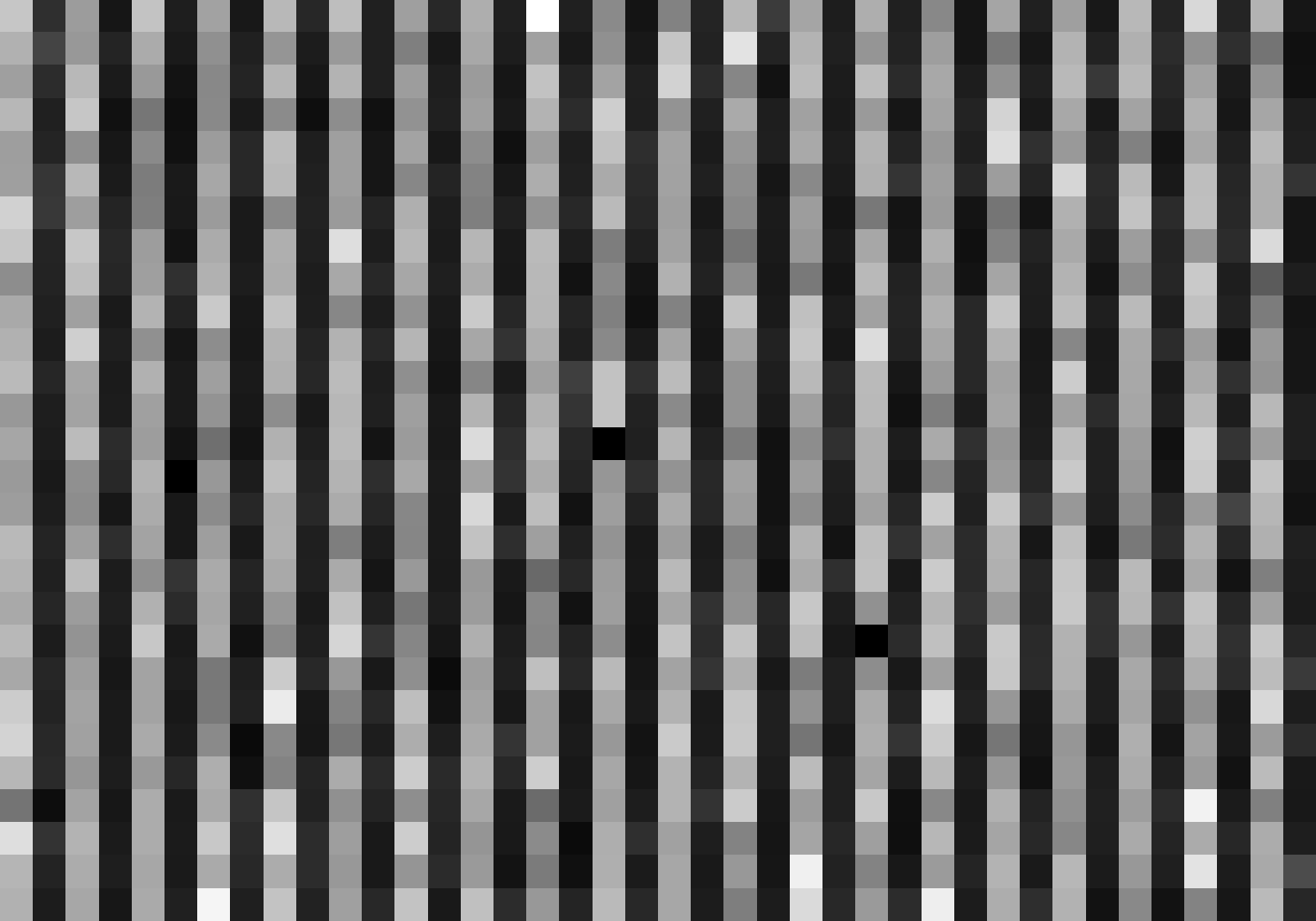}
        \caption{}\label{fig:mask_DF}
    \end{subfigure}
\caption{(a) Diagram of the single-mask dark-field (DF) configuration. The blue strips represent the X-ray beamlets created by the mask, while the squares represent the detector pixels, with their color indicating the X-ray intensity measured by each pixel. (b) Zoomed mask image from the experiment.}\label{fig:DF}
\end{figure}

The second configuration, referred to as dark-field (DF) configuration, is designed to capture dark-field images. The setup is nearly identical to the DPC configuration that we previously showed, in which the mask should be positioned so that the period of the projected mask pattern remains twice the pixel size $p$. The key difference being that the beamlets created by the mask are aligned with the center of every other pixel instead of pixel boundaries, shown in Fig. \ref{fig:DF}. 

In the absence of a sample, this alignment produces an image with high-contrast bright and dark patterns (Fig. \ref{fig:mask_DF}). When a sample is placed in the beam path, some X-ray photons scattered by the sample will strike the dark pixels, decreasing the contrast of the mask pattern locally. Consequently, the dark-field signal can be extracted by comparing the intensity contrast between bright and dark pixels with and without the sample.

The intensity received by each pixel can be derived as:
\begin{equation}
    I_n = [w_e + \alpha_1 (-1)^n] T_n(1-L_n) - \alpha_3 (-1)^n T_n S_n
    \label{eqn:DF}
\end{equation}
where $T_n$ and $L_n$ represent the same signals as in the previous section, and $S_n = z S(x_n)$ represents the average dark-field signal at the corresponding pixel.

Also, the coefficients in the equation ($w_e$, $\alpha_1$, and $\alpha_3$) are parameters related to the mask. $w_e$ is still the equivalent transparency of the mask. Similar to $\alpha$ in Eqn. (\ref{eqn:DPC}), $\alpha_1$ and $\alpha_3$ are related to the contrast of the mask.

From Eqn. (\ref{eqn:DF}), we can see that without the sample ($T_n=1$, $L_n=0$, and $S_n=0$), the image shows a high-contrast mask pattern with a period of two pixels, characterized by the effective transparency $w_e$ and contrast $\alpha_1$. Once the sample is introduced into the beam path, the dark-field signal $S_n$ reduces the local contrast of the mask pattern due to scattering.

The retrieval method can be written as:
\begin{equation}
    \left\{
    \begin{aligned}
        T_n(1-L_n) &= \frac{I_{n}^{(S)}+I_{n+1}^{(S)}}{I_{n}^{(M)}+I_{n+1}^{(M)}}\\
        \frac{\alpha_3}{\alpha_1}S_n =& 1-\frac{I_{n}^{(S)}-I_{n+1}^{(S)}}{I_{n}^{(M)}-I_{n+1}^{(M)}}\cdot\frac{I_{n}^{(M)}+I_{n+1}^{(M)}}{I_{n}^{(S)}+I_{n+1}^{(S)}}
    \end{aligned}\right.
    \label{eqn:DF_retreival}
\end{equation}
Similar to the previous section, the right-hand side of the equations is retrieved attenuation (with Laplacian phase enhancement) and dark field image, respectively. The dark field intensity ranges from zero to one, indication the portion of photons scattered to neighboring pixels. In this configuration, the dark field sensitivity of the imaging system is $\alpha_3/\alpha_1$, related to the masks' transmission function.

\begin{figure}[ht]
	\centering
	\begin{subfigure}{0.4\textwidth}
		\includegraphics[width=\textwidth]{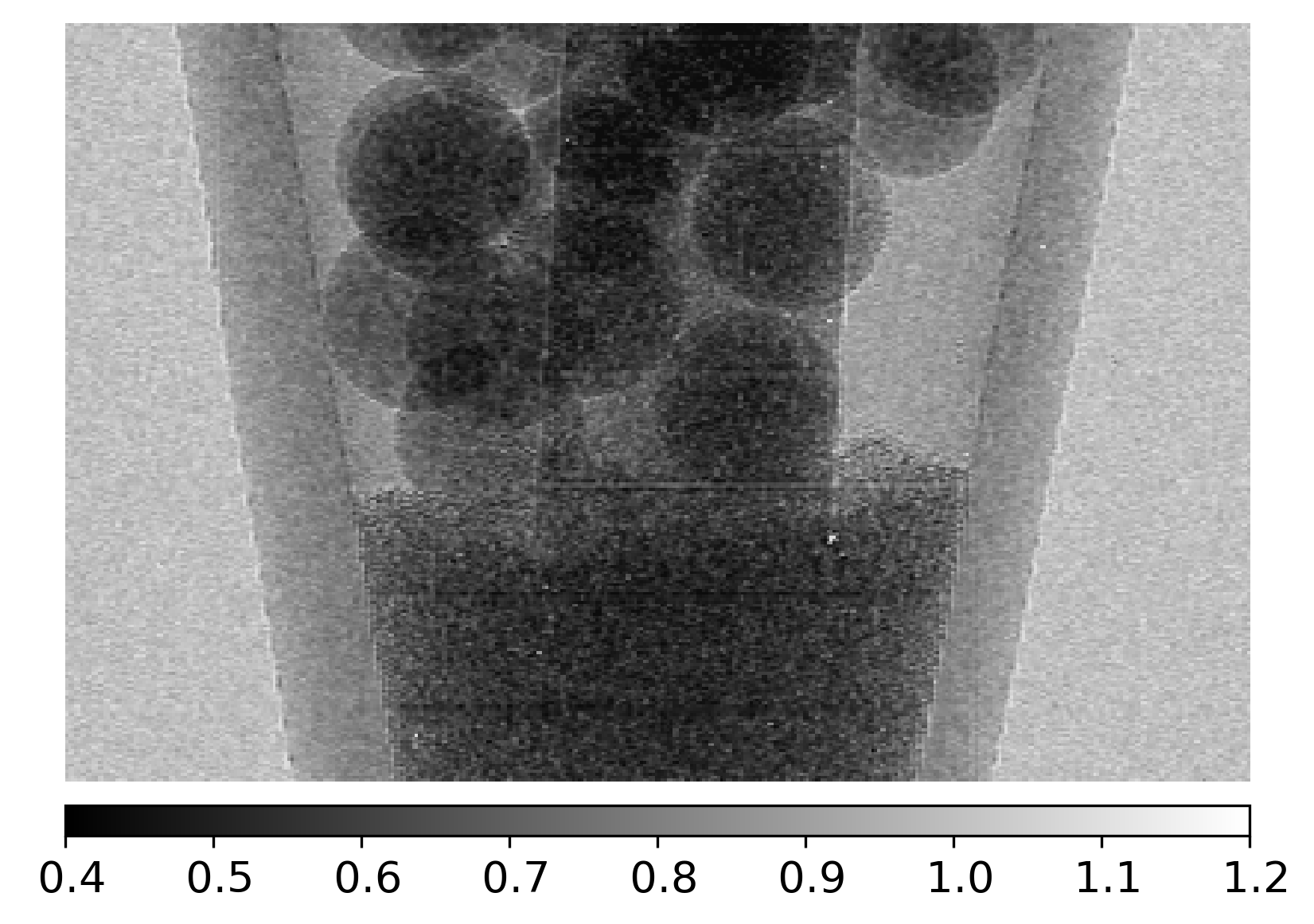}
		\caption{} \label{fig:DF_T}
	\end{subfigure}
	\begin{subfigure}{0.4\textwidth}
		\includegraphics[width=\textwidth]{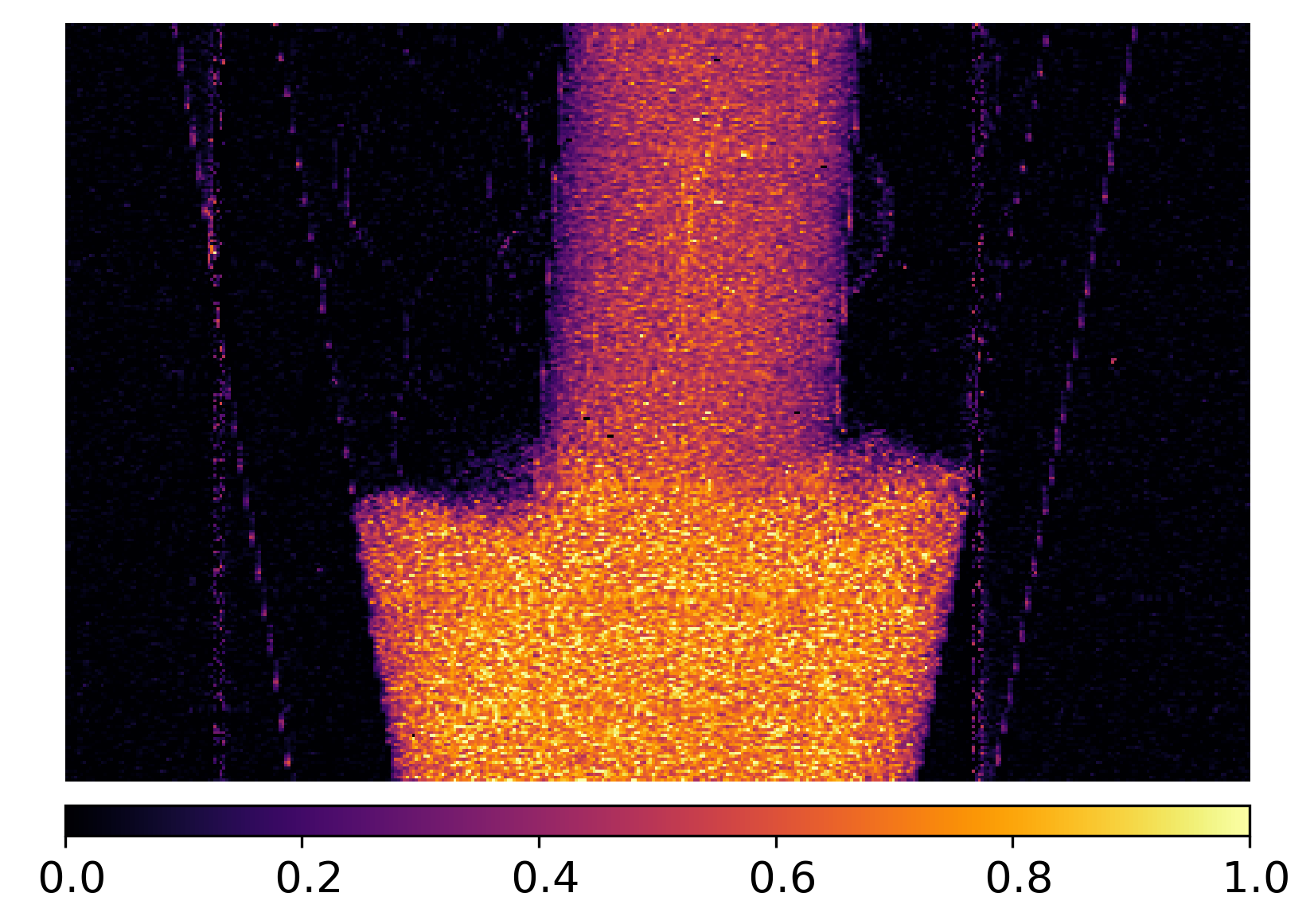}
		\caption{} \label{fig:DF_S}
	\end{subfigure}
	\caption{Retrieved (a) Attenuation image and (b) dark field image of the multi-material phantom with single mask DF configuration}\label{fig:phantom_DF}
\end{figure}

Fig. \ref{fig:phantom_DF} shows the retrieved attenuation and dark-field images obtained from a single shot using the single-mask DF configuration. The sample is the same as the one used in Fig. \ref{fig:phantom_DPC}. From the figure, we can observe that, while the attenuation image (Fig. \ref{fig:DF_T}) is capable of depicting various objects, it falls short in distinguishing between smooth, uniform materials, such as the wall of a centrifuge tube and plastic beads, and materials characterized by microstructures, like graphite and diamond powders. In contrast, the dark field image (Fig. \ref{fig:DF_S}) excels in highlighting materials with microstructures, such as graphite and diamond powders, but show no signal for uniform materials.

However, for capturing differential phase signals, this configuration is ineffective. Since the beamlets are centered on the pixels, micron-level displacements of the beamlet pattern do not alter the total intensity received by each pixel. Thus, this configuration cannot detect differential phase signals.

\subsection{Single Mask DF-DPC Configuration}
\begin{figure} [ht]
    \centering
    \begin{subfigure}{0.5\textwidth}
        \centering
        \includegraphics[width=\textwidth]{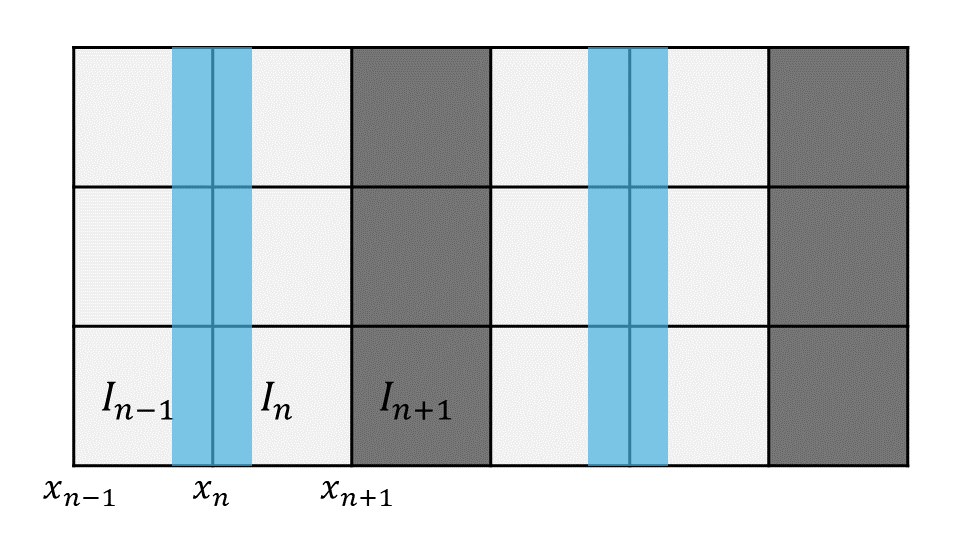}
        \caption{}\label{fig:diagram_DF-DPC}
    \end{subfigure}
    \begin{subfigure}{0.34\textwidth}
        \centering
        \includegraphics[width=\textwidth]{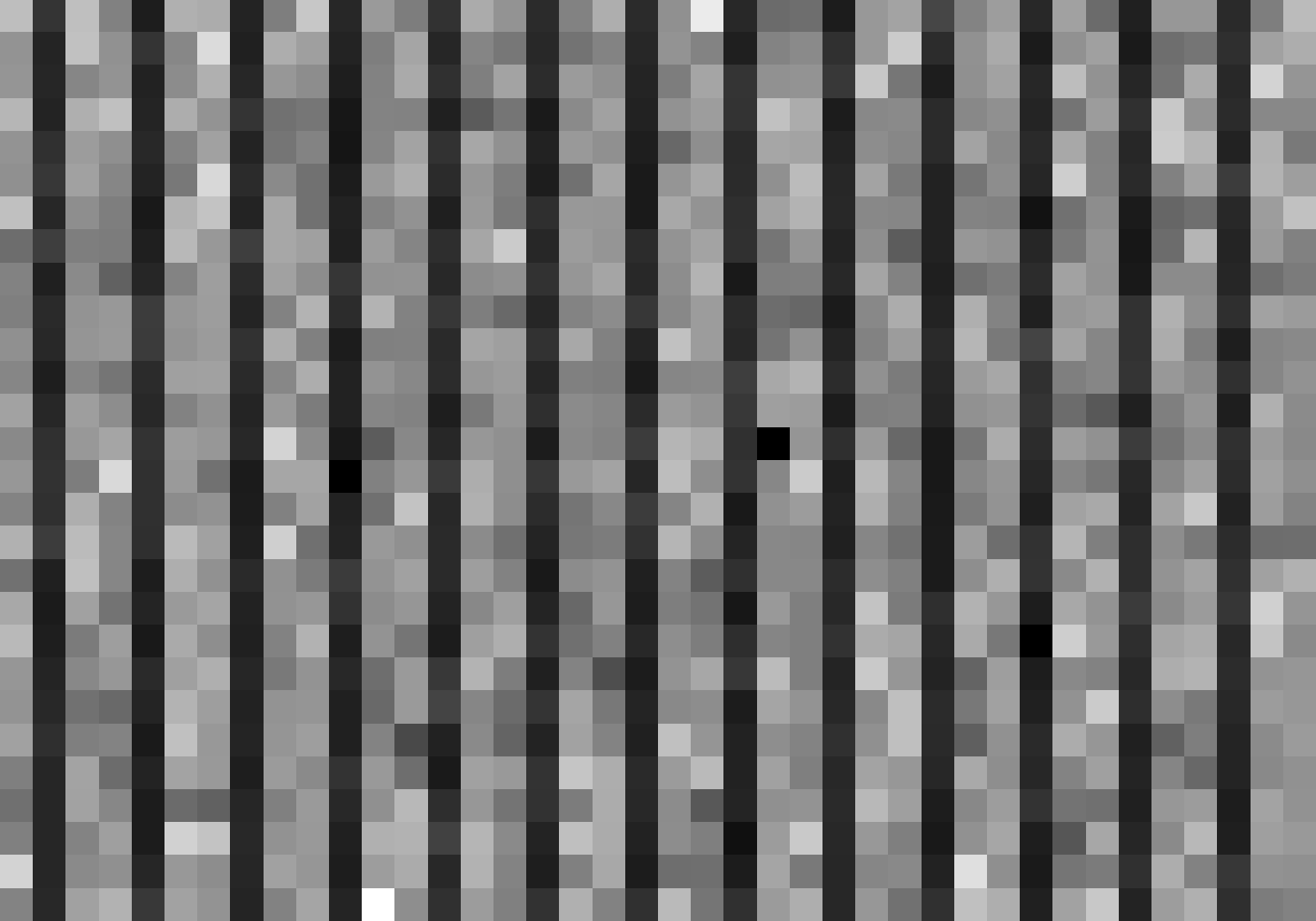}
        \caption{}\label{fig:mask_DF-DPC}
    \end{subfigure}
\caption{(a) Diagram of the single-mask dark-field and differential phase contrast (DF-DPC) configuration. The blue strips represent the X-ray beamlets created by the mask, while the squares represent the detector pixels, with their color indicating the X-ray intensity measured by each pixel. (b) Zoomed mask image from the experiment. This alignment creates a special pattern with three-pixel period.}\label{fig:DF-DPC}
\end{figure}

The third configuration, termed DF-DPC configuration, combines the features of both DPC and DF configurations. In this setup, the mask should be positioned such that the period of the projected mask pattern is three times the detector pixel size $p$, with each beamlet aligned with one in every three-pixel boundaries, as illustrated in Fig. \ref{fig:DF-DPC}. Notably, the same mask used in the first two configurations can be employed here; only its position needs adjustment to achieve the required magnification for this specific arrangement.

As a result, the mask-only image exhibits a pattern with a period of three pixels, where two pixels are equally illuminated (bright pixels) and one pixel remains dark (shown in Fig. \ref{fig:mask_DF-DPC}). This arrangement allows for the extraction of differential phase signals from the two bright pixels, while the dark pixels provide the dark-field signal.

Following the similar process, the measured intensity by each pixel can be derived as:
\begin{equation}
    \left\{
    \begin{aligned}
        I_{n-1} =& (w_e+\frac{1}{2}\alpha_1)T_{n-1}(1-L_{n-1}) + \alpha_2 T_{n-1} D_{n-1} - \frac{1}{2}\alpha_3 T_{n-1} S_{n-1}\\
        I_n =& (w_e+\frac{1}{2}\alpha_1)T_n(1-L_n) - \alpha_2 T_n D_n - \frac{1}{2}\alpha_3 T_n S_n\\
        I_{n+1} =& (w_e-\alpha_1)T_{n+1}(1-L_{n+1}) + \alpha_3 T_{n+1} S_{n+1}
    \end{aligned}\right.
\end{equation}
where $T_n$, $D_n$, $L_n$, $S_n$ represents attenuation, differential phase, Laplacian phase, and dark field signal, respectively. Similar to the previous two configurations, the pixel index $n$ corresponds to the position illustrated in Fig. \ref{fig:diagram_DF-DPC}. 

\begin{figure}[ht]
	\centering
	\begin{subfigure}{0.4\textwidth}
		\includegraphics[width=\textwidth]{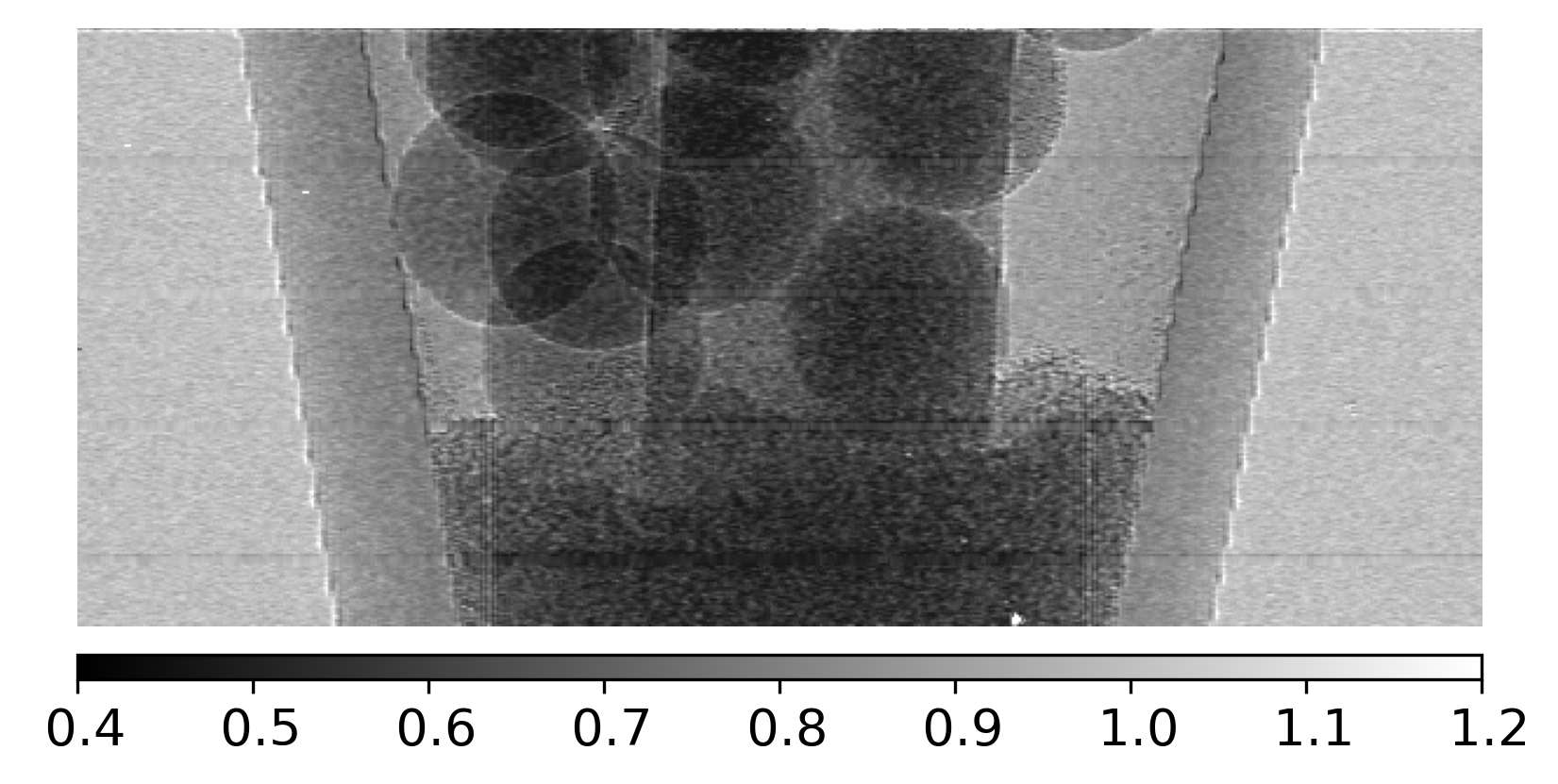}
		\caption{} \label{fig:DF-DPC_T}
	\end{subfigure}
	\begin{subfigure}{0.4\textwidth}
		\includegraphics[width=\textwidth]{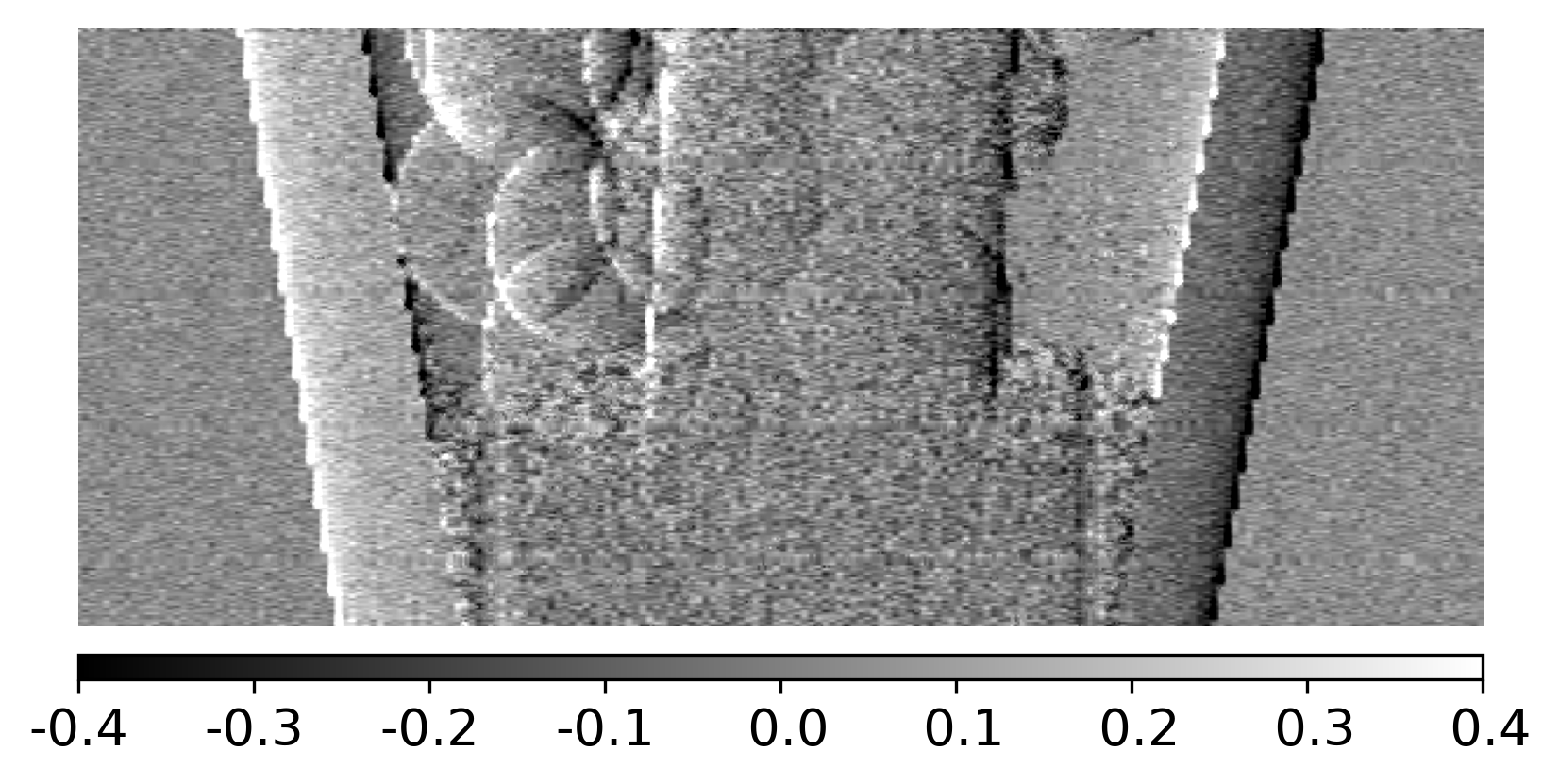}
		\caption{} \label{fig:DF-DPC_D}
	\end{subfigure}
    \begin{subfigure}{0.4\textwidth}
		\includegraphics[width=\textwidth]{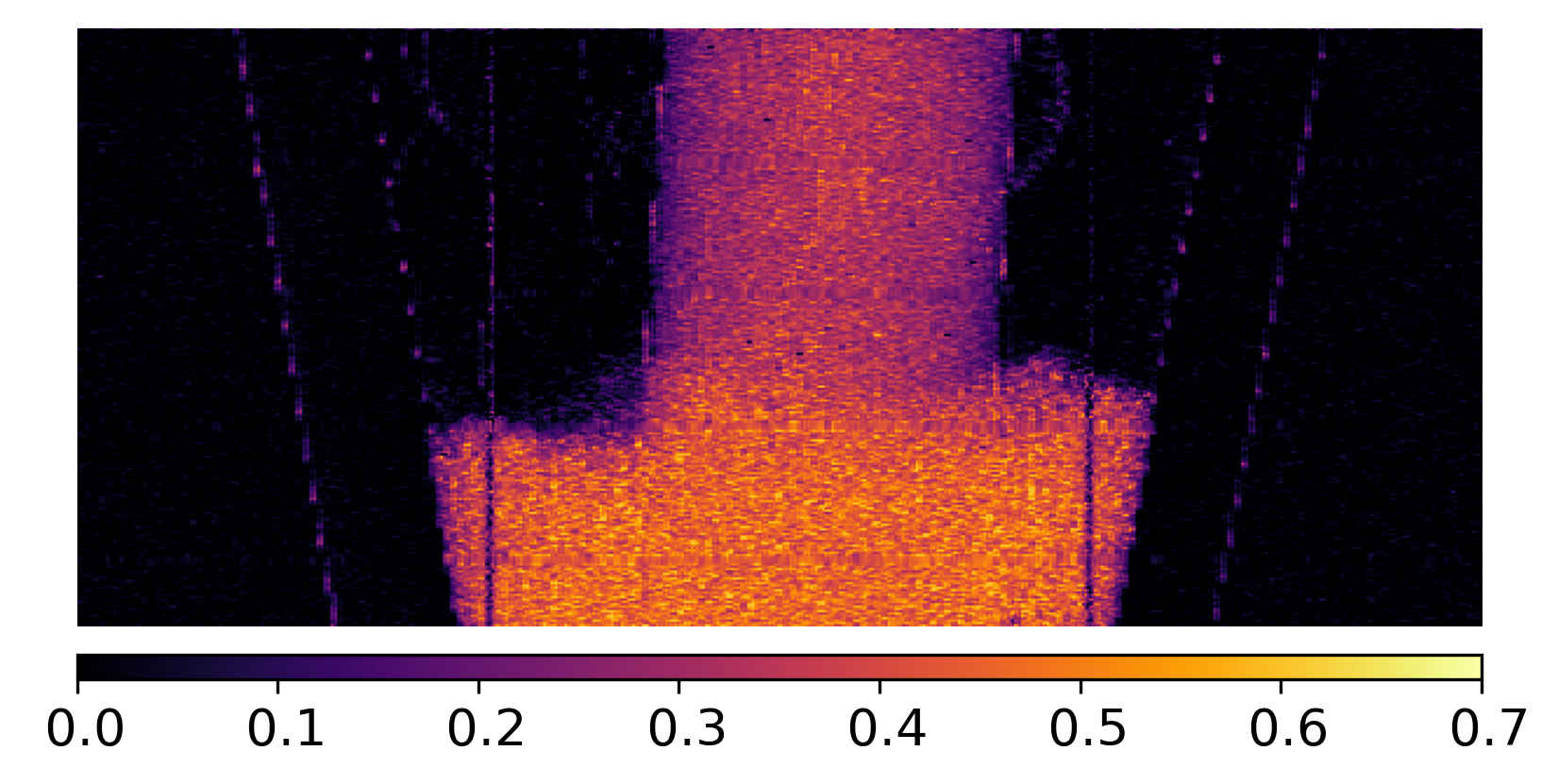}
		\caption{} \label{fig:DF-DPC_S}
	\end{subfigure}
	\caption{Retrieved (a) Attenuation image, (b) DPC image, and (c) dark field image of the multi-material phantom with single mask DF-DPC configuration}\label{fig:phantom_DF-DPC}
\end{figure}

Similar to the previous two configurations, the coefficients in the equation ($w_e$, $\alpha_1$, $\alpha_2$, and $\alpha_3$) are parameters related to the mask. $w_e$ is still the equivalent transparency of the mask. $\alpha_1$, $\alpha_2$, and $\alpha_3$ are similar to $\alpha$, $\alpha_1$ and $\alpha_3$ in Eqn. (\ref{eqn:DPC}) and (\ref{eqn:DF}).

The retrieval method can be derived as:
\begin{equation}
    \left\{
    \begin{aligned}
        T_n(1-L_n&) = \frac{I_{n-1}^{(S)}+I_{n}^{(S)}+I_{n+1}^{(S)}}{I_{n-1}^{(M)}+I_{n}^{(M)}+I_{n+1}^{(M)}}\\
        \frac{2\alpha_2}{w_e+\frac{1}{2}\alpha_1}&D_n = \left(\frac{I_{n-1}^{(S)}}{I_{n-1}^{(M)}} - \frac{I_{n}^{(S)}}{I_{n}^{(M)}}\right)\frac{I_{n-1}^{(M)}+I_{n}^{(M)}+I_{n+1}^{(M)}}{I_{n-1}^{(S)}+I_{n}^{(S)}+I_{n+1}^{(S)}}\\    
        \frac{2\alpha_3}{w_e+2\alpha_1}&S_n = \left[1 - \frac{I_{n-1}^{(S)}+I_{n}^{(S)}-I_{n+1}^{(S)}}{I_{n-1}^{(M)}+I_{n}^{(M)}-I_{n+1}^{(M)}}\cdot\frac{I_{n-1}^{(M)}+I_{n}^{(M)}+I_{n+1}^{(M)}}{I_{n-1}^{(S)}+I_{n}^{(S)}+I_{n+1}^{(S)}}\right]
    \end{aligned}\right.
    \label{eqn:DF-DPC_retreival}
\end{equation}
The right-hand side of the equations is retrieved attenuation, DPC, and dark field image, respectively. The DPC and dark field sensitivity of the imaging system is written as $\frac{2\alpha_2}{w_e+\frac{1}{2}\alpha_1}$ and $\frac{2\alpha_3}{w_e+2\alpha_1}$, respectively.

Fig. \ref{fig:phantom_DF-DPC} shows the retrieved attenuation, differential phase, and dark-field images retrieved from a single shot with the single-mask DF-DPC configuration. The sample is the same as the one used in Fig. \ref{fig:phantom_DPC} and \ref{fig:phantom_DF}. We could see the DF-DPC can produce high-quality attenuation, differential phase, and dark-field images similar to the previous configurations. When comparing configurations, the DF-DPC configuration resulted in higher contrast in DPC images but exhibited lower contrast in dark field images, as opposed to DPC or DF configurations independently.

\section{Results and Discussions}
\begin{figure}[ht]
	\centering
	\begin{subfigure}{0.8\textwidth}
		\includegraphics[width=\textwidth]{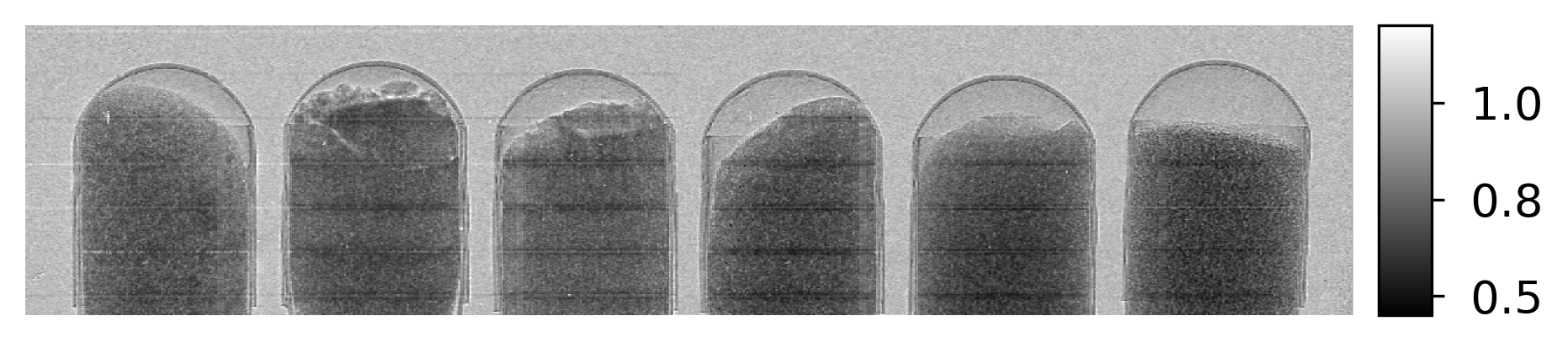}
		\caption{} \label{fig:powder_T}
	\end{subfigure}
	\begin{subfigure}{0.8\textwidth}
		\includegraphics[width=\textwidth]{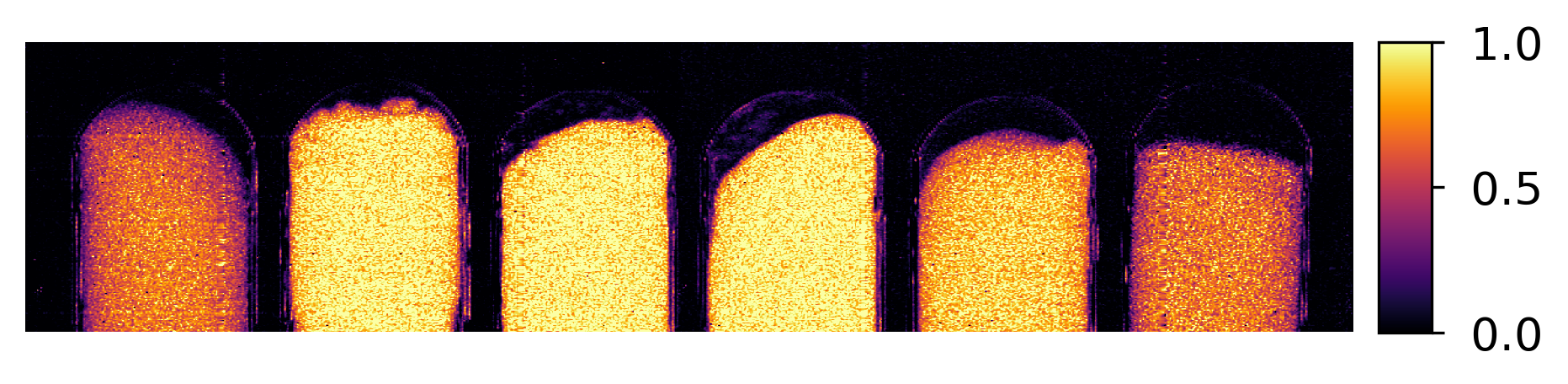}
		\caption{} \label{fig:powder_S}
	\end{subfigure}
	\caption{Retrieved (a) attenuation image and (b) dark field image of six capsules filled with diamond powders of different grain sizes. From left to right corresponds to the sample numbers 1-6 in Table \ref{tab:powder}. Images taken with single-mask DF configuration.}\label{fig:powder}
\end{figure}

\begin{figure}[ht]
	\centering
    \begin{subfigure}{0.46\textwidth}
		\includegraphics[width=\textwidth]{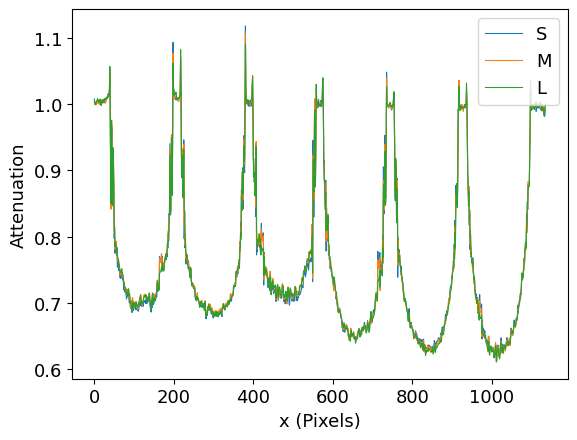}
		\caption{} \label{fig:T_profile_FS}
	\end{subfigure}
    \begin{subfigure}{0.46\textwidth}
		\includegraphics[width=\textwidth]{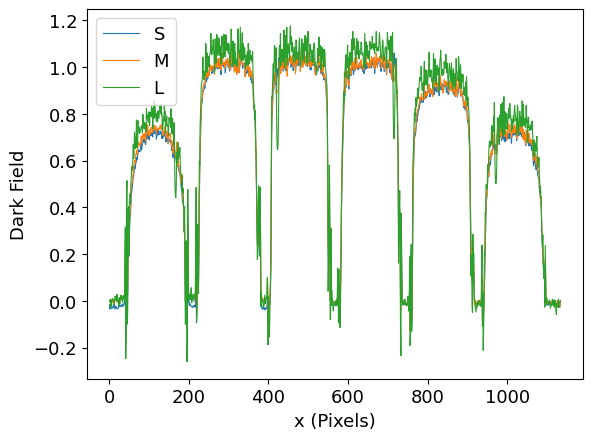}
		\caption{} \label{fig:S_profile_FS}
	\end{subfigure}
    \begin{subfigure}{0.46\textwidth}
		\includegraphics[width=\textwidth]{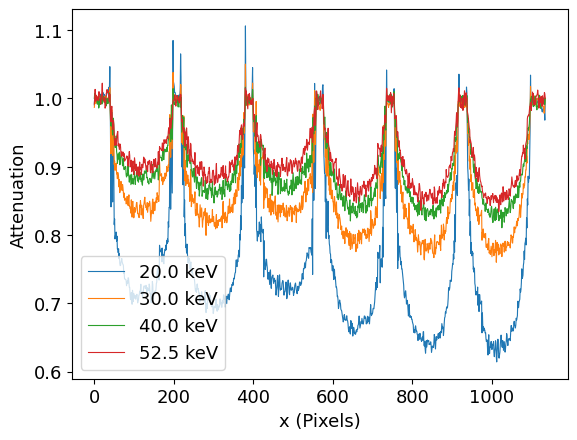}
		\caption{} \label{fig:T_profile_E}
	\end{subfigure}
    \begin{subfigure}{0.46\textwidth}
		\includegraphics[width=\textwidth]{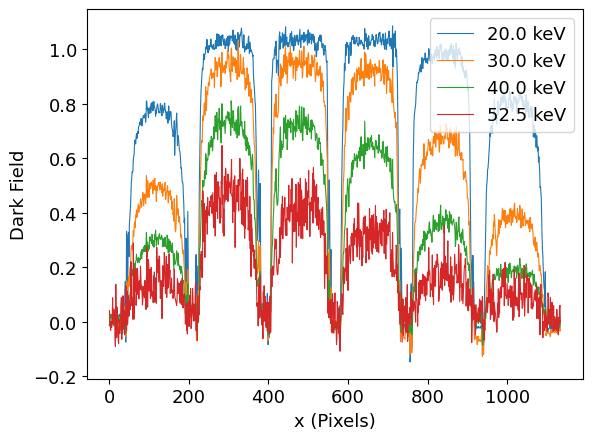}
		\caption{} \label{fig:S_profile_E}
	\end{subfigure}
	\caption{(a)-(b) Cross-sectional profile of retrieved attenuation and dark field image of capsules phantom with different focal spot size of the x-ray source: S (\SI{7}{\micro\metre}), M (\SI{20}{\micro\metre}), and L (\SI{50}{\micro\metre}); (c)-(d) Cross-sectional profile of retrieved attenuation and dark field image of capsules phantom in different energy windows: 15-25 keV, 25-35 keV, 35-45 keV, and 45-60 keV.}\label{fig:profile}
\end{figure}

To investigate the sensitivity of the dark field signal to different feature sizes, we used a phantom consisting of six capsules filled with diamond powders of varying grain sizes. The retrieved attenuation and dark-field images using the single-mask DF configuration are shown in Fig. \ref{fig:powder}. The grain sizes of the diamond powders in each capsule are shown in Table \ref{tab:powder}. The measurements were taken with three different focal spot sizes of the x-ray source (S: \SI{7}{\micro\metre}, M: \SI{20}{\micro\metre}, and L: \SI{50}{\micro\metre}) and across different energy windows (15-25 keV, 25-35 keV, 35-45 keV, and 45-60 keV). The cross-sectional plots are shown in Fig. \ref{fig:profile}.

\begin{table}[h]
\caption{Approximate grain size of each diamond powder sample.}\label{tab:powder}%
\begin{tabular}{cc}
\toprule
Sample Number & Grain Size (\SI{}{\micro\metre})\\
\midrule
1 & 0 - 0.25  \\
2 & 0 - 0.5 \\
3 & 0 - 1 \\
4 & 2 - 3 \\
5 & 10 - 20 \\
6 & 50 - 80 \\ 
\botrule
\end{tabular}
\end{table}

From the attenuation image in Fig. \ref{fig:powder_T}, we can barely see any differences between the six capsules since their infills have almost the same chemical composition. However, the dark field image (Fig. \ref{fig:powder_S}) shows significant differences between each sample. We observe that Samples 2, 3, and 4 exhibit the strongest dark-field signals, indicating that our X-ray dark-field imaging system has the highest sensitivity to features with sizes ranging from 0.25 to \SI{3}{\micro\metre}.

Additionally, Fig. \ref{fig:S_profile_FS} shows that within the tested range, a larger focal spot size does not significantly impact the intensity of the dark field signal. In contrast, according to our previous study \cite{yuanComparingSNRBenefits2022}, the signal intensity of DPC images decreases as focal spot size increases. This suggests that our x-ray dark-field imaging system has a higher potential for real-world applications because it maintains signal intensity regardless of focal spot size, especially in scenarios where maintaining a small focal spot size is challenging.
\begin{figure}[h]
	\centering
	\begin{subfigure}{0.96\textwidth}
		\includegraphics[width=\textwidth]{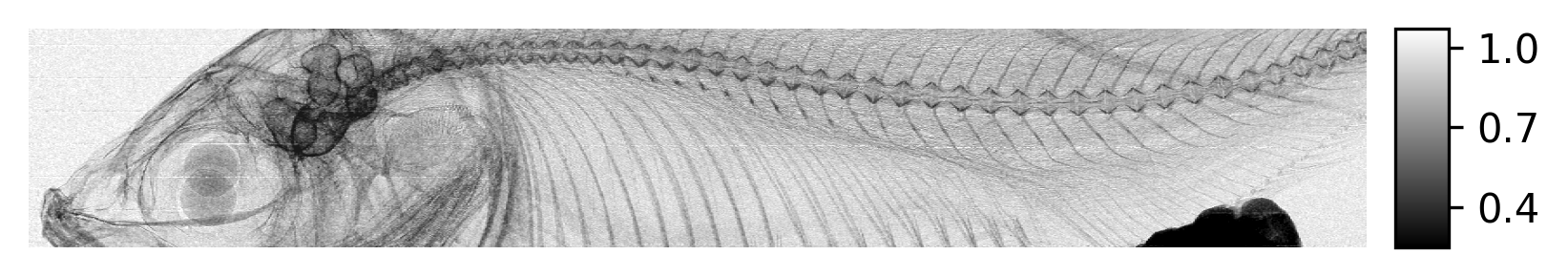}
		\caption{} \label{fig:fish_T}
	\end{subfigure}
	\begin{subfigure}{0.96\textwidth}
		\includegraphics[width=\textwidth]{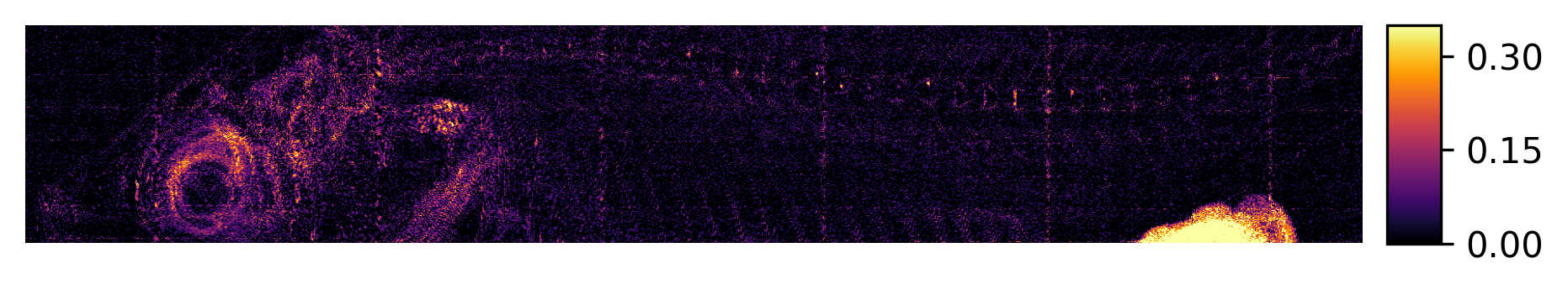}
		\caption{} \label{fig:fish_S}
	\end{subfigure}
    \begin{subfigure}{0.924\textwidth}
		\includegraphics[width=\textwidth]{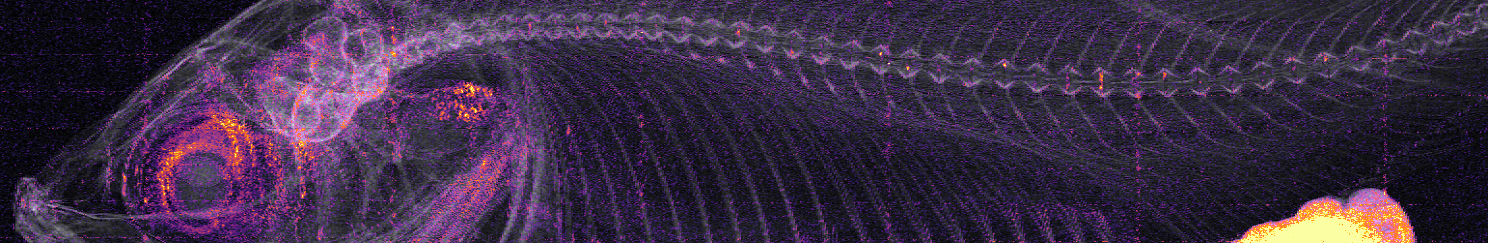}
		\caption{} \label{fig:fish_combined}
	\end{subfigure}
    
	\caption{Retrieved images of a dried fish from single-mask DF configuration. (a) attenuation image; (b) dark field image; (c) image combining attenuation and dark-field signal.}\label{fig:fish}
\end{figure}

Fig. \ref{fig:fish} showcases the retrieved images of a dried fish sample using a single shot with the single-mask DF configuration. The attenuation image (Fig. \ref{fig:fish_T}) provides an overall view of the sample's structure, whereas the dark-field image (Fig. \ref{fig:fish_S}) reveals some special features that are not visible in the attenuation image. The combined image (Fib. \ref{fig:fish_combined}) merges information from both modalities, offering a comprehensive view of the sample, highlighting both structural and microstructural details. These images showed the efficacy of dark-field imaging in capturing detailed structural information that is otherwise undetectable with conventional absorption imaging.

\section{Conclusion}

The single-mask X-ray imaging setup presented in this study represents a significant advancement in multi-contrast imaging, providing a practical and efficient method for simultaneously capturing attenuation, differential phase contrast (DPC), and dark-field images in a single exposure. By eliminating the need for highly coherent X-ray sources, ultra-high-resolution detectors, and intricately fabricated X-ray gratings, our approach simplifies implementation and significantly reduces associated costs, making advanced X-ray imaging more accessible.

The three proposed configurations of the setup offer exceptional flexibility, allowing the imaging approach to be tailored to specific applications—whether emphasizing phase contrast, dark-field signals, or a combination of both. Notably, transitioning between these variations requires no replacement of equipment; only a simple adjustment in the positioning of optical elements is needed. This adaptability enables the achievement of all three imaging modalities within a single system, enhancing its versatility and utility across diverse applications.

Building on our previous work, the light-transport model developed for these three variations provides a clear and intuitive framework for understanding signal formation and offers robust retrieval methods for each contrast type. The model also elucidates the influence of system parameters on the sensitivity of each imaging modality, guiding further optimization and application. Future work will focus on investigating and analyzing the sensitivity of various retrieved features to key design parameters.

Overall, this work lays the groundwork for more accessible and cost-effective X-ray imaging solutions with broad potential applications in clinical diagnostics and industrial inspection. The innovations presented here promise to enhance the capabilities of X-ray imaging systems, making advanced multi-contrast imaging a viable option for a wider range of settings and users.

\section*{Methods}

\noindent\textbf{X-ray Source:}
We utilized a polychromatic micro-focus x-ray tube (Hamamatsu L8121-03) operating with three different focal spot sizes (\SI{7}{\micro\metre}, \SI{20}{\micro\metre}, and \SI{50}{\micro\metre}). The tube voltage we used ranges from 40 to 60 kV. \\

\noindent\textbf{Detector:}
The data was collected using a silicon WidePix photon-counting detector \cite{ballabrigaMedipix3RXHighResolution2013} with a pixel size of \SI{55}{\micro\metre} and sensor thicknes of \SI{500}{\micro\metre}, which was meticulously calibrated and corrected \cite{dasEnergyCalibrationPhoton2015,vespucciRobustEnergyCalibration2019}. The detector sensor measures approximately \SI{70}{\milli\metre} $\times$ \SI{14}{\milli\metre}, consisting of five identical chips tiled together, with each chip containing 256$\times$256 pixels. Spectral information from the photon counting detector was not used for our single shot dark field and differential phase retrievals demonstrated here. The spectral capability of the detector however helped demonstrate the spectral sensitivity to dark field, with results shown in Fig.\ref{fig:profile}. Thus our proposed novel method can work with any standard energy integrating detector enhancing prospect for wider adaptability\\

\backmatter

\bmhead{Acknowledgment}

This work was partially supported by funding from from the NIH National Institute of Biomedical Imaging and Bioengineering (NIBIB) grant R01 EB EB029761, the US Department of Defense (DOD) Congressionally Directed Medical Research Program (CDMRP) Breakthrough Award BC151607 and the National Science Foundation CAREER Award 1652892.


\bibliography{JCLibrary}

\end{document}